\documentclass[12pt]{article}
\usepackage{amssymb}
\begin{document}

%************************** Text Begins here ******************************

%  Greek letters

\def\a{\alpha}
\def\b{\beta}
\def\d{\delta}
\def\e{\epsilon}
\def\g{\gamma}
\def\h{\mathfrak{h}}
\def\k{\kappa}
\def\l{\lambda}
\def\o{\omega}
\def\p{\wp}
\def\r{\rho}
\def\t{\theta}
\def\s{\sigma}
\def\z{\zeta}
\def\x{\xi}
 \def\A{{\cal{A}}}
 \def\B{{\cal{B}}}
 \def\C{{\cal{C}}}
 \def\D{{\cal{D}}}
\def\G{\Gamma}
\def\K{{\cal{K}}}
\def\O{\Omega}
\def\R{\bar{R}}
\def\T{{\cal{T}}}
\def\L{\Lambda}
\def\f{E_{\tau,\eta}(sl_2)}
\def\E{E_{\tau,\eta}(sl_n)}
\def\Zb{\mathbb{Z}}
\def\Cb{\mathbb{C}}

\def\R{\overline{R}}
% Shorthands for \begin{equation} and the like

\def\beq{\begin{equation}}
\def\eeq{\end{equation}}
\def\bea{\begin{eqnarray}}
\def\eea{\end{eqnarray}}
\def\ba{\begin{array}}
\def\ea{\end{array}}
\def\no{\nonumber}
\def\le{\langle}
\def\re{\rangle}
\def\lt{\left}
\def\rt{\right}

\newtheorem{Theorem}{Theorem}
\newtheorem{Definition}{Definition}
\newtheorem{Proposition}{Proposition}
\newtheorem{Lemma}{Lemma}
\newtheorem{Corollary}{Corollary}
\newcommand{\proof}[1]{{\bf Proof. }
        #1\begin{flushright}$\Box$\end{flushright}}

\baselineskip=20pt

%%%%%%%%%%%%%%%%%%%%%%%%%%%%%%%%%%%%%%%%%%%%%%%%%%%%%%%%%%%%
%                                                          %
%  Title page                                              %
%                                                          %
%%%%%%%%%%%%%%%%%%%%%%%%%%%%%%%%%%%%%%%%%%%%%%%%%%%%%%%%%%%%
\newfont{\elevenmib}{cmmib10 scaled\magstep1}
\newcommand{\preprint}{
   \begin{flushleft}
     %\elevenmib Yukawa\, Institute\, Kyoto\\
   \end{flushleft}\vspace{-1.3cm}
   \begin{flushright}\normalsize
  % \sf  YITP-03-53\\
   %  {\tt hep-th/yymmnnn} \\
   July 2006
   \end{flushright}}
\newcommand{\Title}[1]{{\baselineskip=26pt
   \begin{center} \Large \bf #1 \\ \ \\ \end{center}}}
\newcommand{\Author}{\begin{center}
   \large \bf
Wen-Li Yang ${}^{a,b}$,
  Yao-Zhong Zhang ${}^{b,c}$ and Xin Liu ${}^b$\end{center}}
\newcommand{\Address}{\begin{center}

${}^a$ Institute of Modern Physics, Northwest University,
       Xian 710069, P.R. China\\
${}^b$ Department of Mathematics, University of Queensland, Brisbane,
       QLD 4072, Australia\\
${}^c$ Physikalisches Institut, Universit\"at Bonn, D-53115 Bonn, Germany
   \end{center}}
\newcommand{\Accepted}[1]{\begin{center}
   {\large \sf #1}\\ \vspace{1mm}{\small \sf Accepted for Publication}
   \end{center}}

\preprint
\thispagestyle{empty}
\bigskip\bigskip\bigskip

\Title{$gl(4|4)$ current algebra: free field realization and
screening currents} \Author

\Address
\vspace{1cm}

\begin{abstract}
The $gl(4|4)$ current algebra at general level $k$ is
investigated. Its free field representation and
corresponding energy-momentum tensor are constructed. Seven screening
currents of the first kind are also presented.

\vspace{1truecm} \noindent {\it PACS:} 11.25.Hf; 02.20.Tw

\end{abstract}
\newpage
%%%%%%%%%%%%%%%%%%%%%%%%%%%%%%%%%%%%%%%%%%%%%%%%%%%%%%%%%%%%%%%
%                                                             %
%  1. Introduction                                            %
%                                                             %
%%%%%%%%%%%%%%%%%%%%%%%%%%%%%%%%%%%%%%%%%%%%%%%%%%%%%%%%%%%%%%%
\section{Introduction}
\label{intro} \setcounter{equation}{0}

Two-dimensional non-linear sigma models with supermanifold target space
naturally appear in the quantization of superstring
theory on the AdS-type backgrounds. It was shown that the sigma
model on the supergroup $PSU(1,1|2)$ can be used for quantizing
the superstring theory on the $AdS_3\times S^3$ background with
Ramond-Ramond (RR) flux \cite{Ber99,Bers99}. It has been believed
\cite{Met98,Bers99} that the sigma model on $PSU(2,2|4)$ is
related to the string theory on the $AdS_5\times S^5$ background
and that the understanding of the $PSU(2,2|4)$ sigma model (or its
generalization, $GL(4|4)$-WZNW model)  would shed important
lights on the quantization of the string theory on the $AdS_5\times
S^5$ background. The study of the non-linear
sigma model with supermanifold target spaces has also believed to be
relevant for disordered systems and the integer
quantum Hall transition \cite{Ber95,Mud96,Maa97,Bas00, Gur00}.

In most circumstances,  models of interest are believed to be
more complicated than the WZNW models on supergroups. However, even
the WZNW models on supergroups are far from being {\it understood}
\cite{Sch06}. This is largely due to technical reasons
(such as indecomposability of the operator product expansion (OPE),
appearance of logarithms in correlation functions and continuous
modular transformations of the irreducible characters
\cite{Sem03}), combined with the lack of ``physical intuition".

Free field realization \cite{Fra97} has been proved to be a
powerful method in the study the conformal field theory (CFT) such
as WZNW models. In this letter, motivated by the applications
both to superstring theory and condensed matter physics,  we
investigate the $gl(4|4)$ current algebra associated with the $GL(4|4)$
WZNW model at general level $k$.

%%%%%%%%%%%%%%%%%%%%%%%%%%%%%%%%%%%%%%%%%%%%%%%%%%%%%%%%%%%%%%%
%                                                             %
%  2. $gl(4|4)}$  current algebra                             %
%                                                             %
%%%%%%%%%%%%%%%%%%%%%%%%%%%%%%%%%%%%%%%%%%%%%%%%%%%%%%%%%%%%%%%

\section{$gl(4|4)$ current algebra}
\label{XXZ} \setcounter{equation}{0}

Let us start with some basic notation of the $gl(4|4)$ current
algebra, i.e. the $\widehat{gl(4|4)}$ affine superalgebra
\cite{Kac77,Fra96}. Let $\{E_{i,j}|\,i,j=1,\ldots,8\}$  be the
generators of the finite dimensional superalgebra $gl(4|4)$. The
generators obey the (anti-)commutation relations: \bea [E_{i,j},
E_{k,l}]=
\d_{j\,k}E_{i,l}-(-1)^{([i]+[j])([k]+[l])}\d_{i\,l}E_{k,j}.
\label{Def-1}\eea Here and throughout, we adopt the convention:
$[a,b]=ab-(-1)^{[a][b]}ba$. The $\Zb_2$ grading of the generators
is \bea [E_{i,j}]=[i]+[j],\no\eea where
$[1]=\ldots=[4]=0,\,[5]=\ldots=[8]=1$. Due to the non-simplicity
of  $gl(4|4)$, besides the usually quadratic Casimir element \bea
C_1=\sum_{i,j=1}^8\,(-1)^{[j]}E_{i,j}E_{j,i},\label{Casim-1}\eea
there exists another independent quadratic Casimir element \bea
C_2=\sum_{i,j=1}^8\,E_{i,i}E_{j,j}=\lt(\sum_{i=1}^8E_{i,i}\rt)^2.
\label{Casim-2}\eea These two Casimir elements are useful in the
following for the construction of the energy-momentum tensor.

Let $V$ be a $\Zb_2$-graded $(4+4)$-dimensional linear space with
the orthonormal basis $\{|i\rangle,\, i=1,\ldots, 8\}$. The
$\Zb_2$-grading is chosen as:
$[1]=\ldots=[4]=0,\,[5]=\ldots=[8]=1$. In fact $V$ spans the
fundamental representation of $gl(4|4)$, namely, the generators of
the algebra can be realized on $V$ by \bea E_{i,j}=e_{i,j},\quad
i,j=1,\ldots, 8,\no\eea where $e_{i,j}$ is the matrix with entry
$1$ at the $i$-th row and $j$-th column, and zero elsewhere. We
remark that the Casimir element $C_1$ vanishes on $V$, while $C_2$
acts as a multiple of identity on $V$. From the fundamental
representation of $gl(4|4)$, one can define the consistent,
supersymmetric and invariant inner product \cite{Kac77,Fra96} as
follows, \bea \lt(E_{i,j},\,E_{k,l}\rt)=str\lt(e_{i,j}e_{k,l}\rt).
\label{Inner-product}\eea Here $str$ denotes the supertrace, i.e.,
$str(a)=\sum_{i}(-1)^{[i]}\,a_{i\,i}$.

The $gl(4|4)$ current algebra is generated by the currents
$J_{i,j}(z)$ which are associated with the generators $E_{i,j}$ of
$gl(4|4)$. The current algebra at a general level $k$ obeys the
following OPEs \cite{Fra97}, \bea J_{i,j}(z)\,J_{l,m}(w)=
k\frac{\lt(E_{i,j},E_{l,m}\rt)}{(z-w)^2}+\frac{1}{(z-w)}
\lt(\d_{j\,l}\,J_{i,m}(w)-(-1)^{([i]+[j])([l]+[m])}\d_{i\,m}\,
J_{l,j}(w)\rt) .\label{OPE}\eea

%%%%%%%%%%%%%%%%%%%%%%%%%%%%%%%%%%%%%%%%%%%%%%%%%%%%%%%%%%%%%%%
%                                                             %
%  3. Free field realization                                  %
%                                                             %
%%%%%%%%%%%%%%%%%%%%%%%%%%%%%%%%%%%%%%%%%%%%%%%%%%%%%%%%%%%%%%%
\section{Free field realization}
 \label{FFR} \setcounter{equation}{0}

Free field realization of the $gl(4|4)$ currents, {\it in
principle\/}, can be obtained by a general method outlined in
\cite{Fei90,Bou90,Ito90,Fre94,Ras98}, where differential
realizations of the corresponding finite dimensional Lie (super)
algebras play a key role. However, their constructions become very
complicated for higher-rank algebras \cite{Ito90, Ras98, Din03,
Zha05}. We have found a way to obtain explicit expressions of the
differential realizations of $gl(m|n)$. In our approach, the
construction becomes much simpler. In this letter we will restrict
our attention mainly to $gl(4|4)$, as this already illustrates the
main features. Generalizations to other algebras such as
$gl(m|n)$, more details and proofs, will be given  elsewhere
\cite{Yan06}.

Let us introduce $12$ bosonic coordinates
$\{x_{i,j},\,x_{4+i,4+j}|\,1\leq i<j\leq 4\}$ with the
$\Zb_2$-grading: $[x_{i,j}]=0$, and $16$ fermionic coordinates
$\{\t_{i,4+j}|\,i,j=1,\ldots, 4\}$ with the $\Zb_2$-grading:
$[\t_{i,4+j}]=1$. These coordinates satisfy the following
(anti-)commutation relations: \bea &&[x_{i,j},x_{k,l}]=0,\,\,
[\partial_{x_{i,j}},x_{k,l}]=\d_{ik}\d_{jl},\no\\
&&[\t_{i,4+j},\t_{k,4+l}]=0,\,\,
[\partial_{\t_{i,4+j}},\t_{k,4+l}]=\d_{ik}\d_{jl},\no\eea and the
other commutation relations are vanishing. Then the generators
corresponding to the simple roots in the standard (distinguished)
basis \cite{Fra96} can be realized by the following
differential operators:
\bea E_{j,j+1}&=&\sum_{k\leq j-1}x_{k,j}\,\partial_{x_{k,j+1}}+
\partial_{x_{j,j+1}},\,\,1\leq j\leq 3, \label{Diff-1}\\
E_{4,5}&=&\sum_{k\leq 3}x_{k,4}\,\partial_{\t_{k,5}}+\partial_{\t_{4,5}}, \\
E_{4+j,5+j}&=&\sum_{k\leq 4}\t_{k,4+j}\,\partial_{\t_{k,5+j}}+
\sum_{k\leq j-1}x_{4+k,4+j}\,\partial_{x_{4+k,5+j}}+
\partial_{x_{4+j,5+j}},\,\,1\leq j\leq 3,\\
E_{j,j}&=&\sum_{k\leq j-1}x_{k,j}\,\partial_{x_{k,j}}-
\sum_{j+1\leq k\leq 4}x_{j,k}\,\partial_{x_{j,k}} -\sum_{k\leq
4}\t_{j,4+k}\,\partial_{\t_{j,4+k}}+\l_j,\,\,1\leq j\leq 3,\\
E_{4,4}&=&\sum_{k\leq 3}x_{k,4}\,\partial_{x_{k,4}}-\sum_{k\leq
4}\t_{4,4+k}\,\partial_{\t_{4,4+k}}+\l_4,\\
E_{4+j,4+j}&=&\sum_{k\leq 4}\t_{k,4+j}\,\partial_{\t_{k,4+j}}+
\sum_{k\leq j-1}x_{4+k,4+j}\,\partial_{x_{4+k,4+j}}\no\\
&&\qquad\qquad- \sum_{j+1\leq k\leq
4}x_{4+j,4+k}\,\partial_{x_{4+j,4+k}}+\l_{4+j},\,\,1\leq j\leq 4,\\
E_{j+1,j}&=&\sum_{k\leq j-1}
x_{k,j+1}\,\partial_{x_{k,j}}-\sum_{j+2\leq k\leq 4} x_{j,k}\,
\partial_{x_{j+1,k}}-\sum_{k\leq 4}\t_{j,4+k}\,\partial_{\t_{j+1,4+k}} \no\\
&&\quad -x_{j,j+1}\lt(\sum_{j+1\leq k\leq
4}x_{j,k}\,\partial_{x_{j,k}}+\sum_{k\leq 4}
\t_{j,4+k}\,\partial_{\t_{j,4+k}}\rt)\no\\
&&\quad +x_{j,j+1}\lt(\sum_{j+2\leq k\leq
4}x_{j+1,k}\,\partial_{x_{j+1,k}}+\sum_{k\leq
4}\t_{j+1,4+k}\,\partial_{\t_{j+1,4+k}}\rt)\no\\
&&\quad+x_{j,j+1} \lt(\l_j-\l_{j+1}\rt),\,\,1\leq j\leq 3,\\
E_{5,4}&=&\sum_{k\leq 3}\t_{k,5}\,\partial_{x_{k,4}}+\sum_{2\leq
k\leq 4}\t_{4,4+k}\partial_{x_{5,4+k}}\no\\
&&\quad-\t_{4,5}\lt(\sum_{2\leq k\leq
4}\lt(\t_{4,4+k}\,\partial_{\t_{4,4+k}}
+x_{5,4+k}\,\partial_{x_{5,4+k}}\rt)\rt)\no\\
&&\quad +\t_{4,5}\lt(\l_{4}+\l_{5}\rt),\\
E_{5+j,4+j}&=&\sum_{k\leq 4}\t_{k,5+j}\,\partial_{\t_{k,4+j}}
+\sum_{k\leq j-1}x_{4+k,5+j}\,\partial_{x_{4+k,4+j}}-\sum_{j+2\leq
k\leq 4} x_{4+j,4+k}\,\partial_{x_{5+j,4+k}}\no\\
&&\quad -x_{4+j,5+j}\lt(\sum_{j+1\leq k\leq
4}x_{4+j,4+k}\,\partial_{x_{4+j,4+k}}-\sum_{j+2\leq k\leq
4}x_{5+j,4+k}\,\partial_{x_{5+j,4+k}}\rt)\no\\
&&\quad +x_{4+j,5+j}\lt(\l_{4+j}-\l_{5+j}\rt),\,\,1\leq j\leq
3,\label{Diff-2}\eea
where $\{\l_{j}|j=1,\ldots, 8\}$ are $c$-numbers which label the lowest
weight vector of $gl(4|4)$. The other generators associated with
the non-simple roots can be constructed through the simple ones by
the commutation relations,
\bea E_{i,j}&=&[E_{i,k},\,E_{k,j}],\,\,
1\leq i<k<j\leq 8\,\,{\rm and}\,2\leq j-i,\label{Non-simple-1}\\
E_{j,i}&=&[E_{j,k},\,E_{k,i}],\,\, 1\leq i<k<j\leq 8 \,\,{\rm
and}\,2\leq j-i.\label{Non-simple-2} \eea One can prove that the
differential realization (\ref{Diff-1})-(\ref{Diff-2}) of
$gl(4|4)$ satisfies the commutation relations (\ref{Def-1}).

With the help of the differential realization we can obtain the
free field realization (Wakimoto construction) of the $gl(4|4)$
current algebra in terms of $12$ bosonic $\b$-$\g$ pairs
($(\b_{i,j},\,\g_{i,j})$ and $(\bar{\b}_{i,j}\bar{\g}_{i,j})$, for
$1\leq i<j\leq 4$), $16$ fermionic $b$-$c$ pairs
(($\psi^{\dagger}_{i,j},\,\psi_{i,j})$, for $1\leq i,j\leq 4$) and
$8$ free scalar fields ($\phi_i$, for $i=1,\ldots,8$). The free
fields obey the following OPEs:\bea
\b_{i,j}(z)\,\g_{k,l}(w)&=&-\g_{k,l}(z)\,\b_{i,j}(w)=
\frac{\d_{ik}\d_{jl}}{(z-w)},\,\,1\leq i<j\leq 4\,{\rm and}\,1\leq
k<l\leq
4,\label{OPE-F-1}\\
\bar{\b}_{i,j}(z)\,\bar{\g}_{k,l}(w)&=&-\bar{\g}_{k,l}(z)\,
\bar{\b}_{i,j}(w)= \frac{\d_{ik}\d_{jl}}{(z-w)},\,\,1\leq i<j\leq
4\,{\rm and}\,1\leq k<l\leq
4,\\
\psi_{i,j}(z)\psi_{k,l}^{\dagger}(w)&=&\psi_{k,l}^{\dagger}(z)
\psi_{i,j}(w)=\frac{\d_{ik}\d_{jl}}{(z-w)},\,\,\,\,\,\,1\leq
i,j\leq 4\,\,{\rm and}\,\,1\leq k,l\leq
4,\\
\phi_i(z)\phi_j(w)&=&(-1)^{[i]}\,\d_{ij}\,\ln(z-w),\,\,\,\,\,1\leq
i,j\leq 8,\label{OPE-F-2}\eea  and the other OPEs are trivial.

The free field realization of the $gl(4|4)$ current algebra
(\ref{OPE}) is obtained by the following substitution: \bea
&&x_{i,j}\longrightarrow \g_{i,j}(z),\quad \partial_{x_{i,j}}
\longrightarrow \b_{i,j}(z),\quad 1\leq i<j\leq 4,\no\\
&&x_{4+i,4+j}\longrightarrow \bar{\g}_{i,j}(z),\quad
\partial_{x_{4+i,4+j}} \longrightarrow \bar{\b}_{i,j}(z),
\quad 1\leq i<j\leq 4,\no\\
&&\t_{i,4+j}\longrightarrow \psi^{\dagger}_{i,j}(z),\quad
\partial_{\t_{i,4+j}} \longrightarrow \psi_{i,j}(z),\quad 1\leq i,j\leq
4,\no\\
&&\l_j\longrightarrow \sqrt{k}\partial\phi_j(z)-
\frac{(-1)^{[j]}}{2\sqrt{k}}\sum_{l=1}^8\phi_l(z),\,\, 1\leq j\leq
8,\no\eea in the differential realization
(\ref{Diff-1})-(\ref{Diff-2}) of $gl(4|4)$ and a subsequent
addition of anomalous terms linear in $\partial
\psi^{\dagger}(z)$, $\partial\g(z)$ and $\partial\bar{\g}(z)$ in
the expressions of the currents. Here we present the
realization of the currents associated with the simple roots, \bea
J_{j,j+1}(z)&=&\sum_{l\leq
j-1}\g_{l,j}(z)\b_{l,j+1}(z)+\b_{j,j+1}(z),\,\,1\leq j\leq
3,\label{Fre-currents-1}\\
J_{4,5}(z)&=&\sum_{l\leq
3}\g_{l,4}(z)\psi_{l,1}(z)+\psi_{4,1}(z),\\
J_{4+j,5+j}(z)&=&\sum_{l\leq
4}\psi^{\dagger}_{l,j}(z)\psi_{l,j+1}(z)+\sum_{l\leq
j-1}\bar{\g}_{l,j}(z)\bar{\b}_{l,j+1}(z)+\bar{\b}_{j,j+1}(z),\,\,
1\leq j\leq 3,\\
J_{j,j}(z)&=&\sum_{l\leq j-1}\g_{l,j}(z)\b_{l,j}(z)- \sum_{j+1\leq
l\leq 4}\g_{j,l}(z)\b_{j,l}(z) -\sum_{l\leq
4}\psi^{\dagger}_{j,l}(z)\psi_{j,l}(z)\no\\
&&\quad+\sqrt{k}\partial\phi_j(z)-\frac{1}{2\sqrt{k}}
\sum_{l=1}^8\partial\phi_l(z),\,\,1\leq j\leq 3,\\
J_{4,4}(z)&=&\hspace{-0.1truecm}\sum_{l\leq
3}\g_{l,4}(z)\b_{l,4}(z)-\sum_{l\leq
4}\psi^{\dagger}_{4,l}(z)\psi_{4,l}(z)+ \sqrt{k}\partial\phi_4(z)
\no\\&&\quad -\frac{1}{2\sqrt{k}}
\sum_{l=1}^8\partial\phi_l(z),\\
J_{4+j,4+j}(z)&=&\sum_{l\leq
4}\psi^{\dagger}_{l,j}(z)\psi_{l,j}(z)+ \sum_{l\leq
j-1}\bar{\g}_{l,j}(z)\bar{\b}_{l,j}(z)- \sum_{j+1\leq l\leq
4}\bar{\g}_{j,l}(z)\bar{\b}_{j,l}(z)\no\\
&&\quad+\sqrt{k}\partial\phi_{4+j}(z)+\frac{1}{2\sqrt{k}}
\sum_{l=1}^8\partial\phi_l(z),\,\,1\leq j\leq 4, \\
J_{j+1,j}(z)&=&\sum_{l\leq j-1}
\g_{l,j+1}(z)\b_{l,j}(z)-\sum_{j+2\leq l\leq 4} \g_{j,l}(z)
\b_{j+1,l}(z)-\sum_{l\leq 4}\psi^{\dagger}_{j,l}(z)\psi_{j+1,l}(z)
\no\\
&&\quad -\g_{j,j+1}(z)\lt(\sum_{j+1\leq l\leq
4}\g_{j,l}(z)\b_{j,l}(z)+\sum_{l\leq 4}
\psi^{\dagger}_{j,l}(z)\psi_{j,l}(z)\rt)\no\\
&&\quad +\g_{j,j+1}(z)\lt(\sum_{j+2\leq l\leq
4}\g_{j+1,l}(z)\b_{j+1,l}(z)+\sum_{l\leq
4}\psi^{\dagger}_{j+1,l}(z)\psi_{j+1,l}(z)\rt)\no\\
&&\quad+\sqrt{k}\g_{j,j+1}(z)
\lt(\partial\phi_j(z)-\partial\phi_{j+1}(z)\rt)
+(k+j-1)\partial\g_{j,j+1}(z),\no\\
&&\quad\qquad\,\,1\leq j\leq 3,\\
J_{5,4}(z)&=&\sum_{l\leq
3}\psi^{\dagger}_{l,1}(z)\b_{l,4}(z)+\sum_{2\leq
l\leq 4}\psi^{\dagger}_{4,l}(z)\bar{\b}_{1,l}(z)\no\\
&&\quad-\psi^{\dagger}_{4,1}(z)\lt(\sum_{2\leq l\leq
4}\lt(\psi^{\dagger}_{4,l}(z)\,\psi_{4,l}(z)
+\bar{\g}_{1,l}(z)\bar{\b}_{1,l}(z)\rt)\rt)\no\\
&&\quad +\sqrt{k}\psi^{\dagger}_{4,1}(z)
\lt(\partial\phi_{4}(z)+\partial\phi_{5}(z)\rt)
+(k+3)\partial\psi^{\dagger}_{4,1}(z),\\
J_{5+j,4+j}(z)&=&\sum_{l\leq
4}\psi^{\dagger}_{l,j+1}(z)\psi_{l,j}(z) +\sum_{l\leq
j-1}\bar{\g}_{l,j+1}(z)\bar{\b}_{l,j}(z)-\sum_{j+2\leq
l\leq 4} \bar{\g}_{j,l}(z)\bar{\b}_{j+1,l}(z)\no\\
&&\quad -\bar{\g}_{j,j+1}(z)\lt(\sum_{j+1\leq l\leq
4}\bar{\g}_{j,l}(z)\bar{\b}_{j,l}(z)-\sum_{j+2\leq l\leq
4}\bar{\g}_{j+1,l}(z)\bar{\b}_{j+1,l}(z)\rt)\no\\
&&\quad +\sqrt{k}\bar{\g}_{j,j+1}
\lt(\partial\phi_{4+j}(z)-\partial\phi_{5+j}(z)\rt)
-(k+5-j)\partial\bar{\g}_{j,j+1}(z),\no\\
&&\qquad\qquad\,\,1\leq j\leq 3.\label{Fre-currents-2}\eea Here
normal ordering of the free field expressions is implied. The
free field realization for other currents associated with the
non-simple roots can be obtained from the OPEs of the simple ones.
It is straightforward
to check that the above free field realization of the currents
satisfy the OPEs of the $gl(4|4)$ current algebra given in the last section.

%%%%%%%%%%%%%%%%%%%%%%%%%%%%%%%%%%%%%%%%%%%%%%%%%%%%%%%%%%%%%%%
%                                                             %
%  4. Energy-momentum tensor                                  %
%                                                             %
%%%%%%%%%%%%%%%%%%%%%%%%%%%%%%%%%%%%%%%%%%%%%%%%%%%%%%%%%%%%%%%

\section{Energy-momentum tensor}
\label{EMT} \setcounter{equation}{0}

In order to apply the free field realization of the $gl(4|4)$ currents
to compute conformal blocks of the associated WZNW-conformal field
theory, we need to calculate the energy-momentum tensor of the
associated CFT. The Sugawara tensor corresponding to the quadratic
Casimir $C_1$ is given by \bea
T_1(z)&=&\frac{1}{2k}\sum_{i,j=1}^8(-1)^{[j]}:J_{i,j}(z)J_{j,i}(z):\no\\
&=&\frac{1}{2}\sum_{l=1}^8(-1)^{[l]}\partial\phi_l(z)\partial\phi_l(z)
+\frac{1}{2\sqrt{k}}\partial^2
\lt(\sum_{i=1}^4(2i-1)\lt(\phi_i(z)+\phi_{9-i}(z)\rt)\rt)\no\\
&&\quad+\sum_{1=i<j}^4\lt(\partial\g_{i,j}(z)\b_{i,j}(z)
+\partial\bar{\g}_{i,j}(z)\bar{\b}_{i,j}(z)\rt)
+\sum_{i,j=1}^4\partial\psi^{\dagger}_{i,j}(z)\psi_{i,j}(z)\no\\
&&\quad-\frac{1}{2k}\partial\bar{\phi}(z)\,\partial\bar{\phi}(z),\eea
where $\bar{\phi}(z)=\sum_{l=1}^8\phi_l(z)$ and the normal
ordering of the free field expressions is implicit. On the other
hand,  the Sugawara tensor corresponding to the
quadratic Casimir $C_2$ is given by \bea
T_2(z)&=&\frac{1}{2k}\sum_{i,j=1}^8:J_{i,i}(z)J_{j,j}(z):
=\frac{1}{2}:\partial\bar{\phi}(z)\,\partial\bar{\phi}(z):.\eea In
order that all currents $J_{i,j}(z)$ are primary fields with
conformal dimensional one, we define the energy-momentum tensor
$T(z)$ as follow: \bea T(z)&=&T_1(z)+\frac{1}{k}T_2(z)\no\\
&=&\frac{1}{2}\sum_{l=1}^8(-1)^{[l]}\partial\phi_l(z)\partial\phi_l(z)
+\frac{1}{2\sqrt{k}}\partial^2
\lt(\sum_{i=1}^4(2i-1)\lt(\phi_i(z)+\phi_{9-i}(z)\rt)\rt)\no\\
&&\quad+\sum_{1=i<j}^4\lt(\partial\g_{i,j}(z)\b_{i,j}(z)
+\partial\bar{\g}_{i,j}(z)\bar{\b}_{i,j}(z)\rt)
+\sum_{i,j=1}^4\partial\psi^{\dagger}_{i,j}(z)\psi_{i,j}(z),
\label{Energy-Momentum}\eea
where the normal ordering of the free field expressions is
implicit. We find \bea
T(z)T(w)=\frac{c/2}{(z-w)^4}+\frac{2T(w)}{(z-w)^2}+\frac{\partial
T(w)}{(z-w)},\eea with a center charge $c=0$. Moreover, it is easy
to check that \bea
T(z)J_{i,j}(w)=\frac{J_{i,j}(w)}{(z-w)^2}+\frac{\partial
J_{i,j}(w)}{(z-w)},\,\,1\leq i,j\leq 8.\eea Therefore, $T(z)$ is
the energy-momentum tensor of the $gl(4|4)$ current algebra.

%%%%%%%%%%%%%%%%%%%%%%%%%%%%%%%%%%%%%%%%%%%%%%%%%%%%%%%%%%%%%%%
%                                                             %
%  5. Screening currents                                      %
%                                                             %
%%%%%%%%%%%%%%%%%%%%%%%%%%%%%%%%%%%%%%%%%%%%%%%%%%%%%%%%%%%%%%%

\section{Screening currents}
\label{SC}  \setcounter{equation}{0}

An important object in applying the free field realization to the
computation of correlation functions  of the associated CFT is
screening currents. A screening current is a primary field with
conformal dimension one and has the property that the singular
part of the OPE with the affine currents is a total derivative.
These properties ensure that integrated screening currents
(screening charges) may be inserted into correlators while the
conformal or affine Ward identities remain intact. This in turn
makes them very useful in computation of the correlation functions
\cite{Dos84,Ber90}. For the present case, we find seven screening
currents \bea S_j(z)&=&\lt(\sum_{j+2\leq l\leq
4}\g_{j+1,l}(z)\b_{j,l}(z)
+\sum_{l=1}^4\psi^{\dagger}_{j+1,l}(z)\psi_{j,l}(z)+\b_{j,j+1}(z)\rt)
\tilde{s}_j(z),\no\\
&&\qquad\qquad\,\, 1\leq j\leq 3,\label{Screening-1}\\
S_4(z)&=&\lt(\sum_{2\leq
l}\bar{\g}_{1,l}(z)\psi_{4,l}(z)+\psi_{4,1}(z)\rt)\tilde{s}_4(z),\\
S_{4+j}(z)&=&\lt(\sum_{j+2\leq l\leq
4}\bar{\g}_{j+1,l}(z)\bar{\b}_{j,l}(z)+\bar{\b}_{j,j+1}(z)\rt)
\tilde{s}_{4+j}(z) ,\,\,1\leq j\leq 3,\eea where \bea
\tilde{s}_j(z)=e^{-\frac{1}{\sqrt{k}}(\phi_j(z)-\phi_{j+1}(z))},\,\,
\tilde{s}_4(z)=e^{-\frac{1}{\sqrt{k}}(\phi_{4}(z)+\phi_{5}(z))},\,\,
\tilde{s}_{4+j}(z)=e^{\frac{1}{\sqrt{k}}(\phi_{4+j}(z)-\phi_{5+j}(z))}.
\label{Screening-2}\eea The normal ordering of the free field
expressions is implicit in the above equations. The nontrivial
OPEs of the screening currents with the energy-momentum tensor and
the $gl(4|4)$ currents
(\ref{Fre-currents-1})-(\ref{Fre-currents-2}) are \bea &&
T(z)S_j(w)=\frac{S_j(w)}{(z-w)^2}+\frac{\partial S_j(w)}{(z-w)}
=\partial_w\lt\{\frac{S_j(w)}{(z-w)}\rt\},\,\,1\leq j\leq 7,\\
&&J_{i+1,i}(z)S_j(w)=(-1)^{[i]+[i+1]}\d_{ij}\,
\partial_{w}\lt\{\frac{k \,\tilde{s}_j(w)}{(z-w)}\rt\},\,\,
1\leq i,j\leq 7. \eea The screening currents obtained here are
associated with the simple roots and correspond to the first kind
screening currents \cite{Ber86}. Moreover, $S_4(z)$ is fermionic
screening current and the others are all bosonic ones.

%%%%%%%%%%%%%%%%%%%%%%%%%%%%%%%%%%%%%%%%%%%%%%%%%%%%%%%%%%%%%%%
%                                                             %
%  6. Discussions                                             %
%                                                             %
%%%%%%%%%%%%%%%%%%%%%%%%%%%%%%%%%%%%%%%%%%%%%%%%%%%%%%%%%%%%%%%

\section{Discussions}
\label{Con} \setcounter{equation}{0}

We have studied the $gl(4|4)$ current algebra at general level
$k$. We have constructed its Wakimoto free field realization
(\ref{Fre-currents-1})-(\ref{Fre-currents-2}) and the
corresponding  energy-momentum tensor (\ref{Energy-Momentum}).
We have also found seven screening currents,
(\ref{Screening-1})-(\ref{Screening-2}), of the first kind.

To fully take the advantage of the CFT method, one needs to
construct its primary fields. It is well-known that there exist
two types of representations for the underlying finite dimensional
superalgebra $gl(4|4)$: typical and atypical representations.
Atypical representations have no counterpart in the bosonic
algebra setting and the understanding of such representations is still
very much incomplete. Although the construction of the primary
fields associated with typical representations are similar to
the bosonic algebra cases, it is a highly non-trivial
task to construct the primary fields associated with atypical
representations \cite{Zha05}.

%%%%%%%%%%%%%%%%%%%%%%%%%%%%%%%%%%%%%%%%%%%%%%%%%%%%%%%%%%%%%%%
%                                                             %
%  Acknowledgments                                            %
%                                                             %
%%%%%%%%%%%%%%%%%%%%%%%%%%%%%%%%%%%%%%%%%%%%%%%%%%%%%%%%%%%%%%%
\section*{Acknowledgements}
The financial support from  the Australian Research Council is
gratefully acknowledged. YZZ was also supported by the Max-Planck-Institut f\"ur
Mathematik (Bonn) and by the Alexander von Humboldt-Stiftung.
XL has been supported by IPRS and UQGSS scholarships of
the University of Queensland. YZZ would like to thank the Max-Planck-Institut
f\"ur Mathematik, where part of this work was done, and Physikalisches
Institut der Universit\"at Bonn, especially G\"unter von Gehlen, for
hospitality.

\vspace{1.00truecm}

\noindent{{\large \it Note added}}: \, We became aware that free
field realizations of $sl(N|N)$ (or $gl(N|N)$) current algebra
were investigated previously in \cite{Bar91,Isi94}. There, the
$sl(N|N)_k$ (or $gl(N|N)_k$) currents were expressed in terms of
the $sl(N)_{k-N}$ and $sl(N)_{-k-N}$ currents and some $b$-$c$
pairs. As part of the results of our paper, we give the {\em
explicit} expressions of $gl(4|4)_k$ currents in terms of free
fields, by using a different method.

%%%%%%%%%%%%%%%%%%%%%%%%%%%%%%%%%%%%%%%%%%%%%%%%%%%%%%%%%%%%%%%
%                                                             %
%  References                                                 %
%                                                             %
%%%%%%%%%%%%%%%%%%%%%%%%%%%%%%%%%%%%%%%%%%%%%%%%%%%%%%%%%%%%%%%

\end{document}